\documentclass{article}

\usepackage{graphicx} 

\usepackage[english]{babel}

\DeclareUnicodeCharacter{202F}{\,} 

\usepackage{natbib}
\usepackage{csquotes}

\usepackage[normalem]{ulem}
\useunder{\uline}{\ul}{}

\usepackage{booktabs}
\usepackage{tablefootnote} 
\usepackage{threeparttable} 
\usepackage{multirow}
\usepackage{array}
\usepackage{booktabs}
\usepackage{longtable}
\usepackage{lscape}
\usepackage{makecell}

\usepackage{makecell}

\usepackage{tabularx}

\usepackage{hyperref}

\usepackage{xcolor}

\usepackage{comment}

\newcounter{example}[section]

\usepackage{ dsfont }

\providecommand{\keywords}[1]
{
  \small	
  \textbf{\textit{Keywords---}} #1
}

\newcommand{\fre}[1]{#1}

\newcommand{\verberg}[1]{}
\newcommand{\hide}[1]{}
\newcommand{\platform}[1]{\emph{#1}}   

\title{Sorting Methods for Online Deliberation: \\ Towards a Principled Approach}

\author{Nicolien Janssens$^{1}$, Frederik Van De Putte$^{1}$  \\
        \small $^{1}$ Erasmus Institute for Philosophy \& Economics (EIPE), \fre{Erasmus School of Philosophy}, \\
        \small Erasmus University Rotterdam  \\
}

\date{}

\begin{document}

\maketitle

\begin{abstract}
Recent years have seen an increase in the use of online deliberation platforms (DPs). One of the main objectives of DPs is to enhance democratic participation, by allowing citizens to post, comment, and vote on policy proposals. But in what order should these proposals be listed? This paper makes a start with the principled evaluation of \emph{sorting methods} on DPs. First, we introduce a conceptual framework that allows us to classify and compare sorting methods in terms of their purpose and the parameters they take into account. Second, we observe that the choice for a sorting method is often \emph{ad hoc} and rarely justified. Third and last, we criticise sorting by number of approvals (``likes''), a method that is very common in practice. On the one hand, we show that \emph{if} approvals are used for sorting, this should be done in an integrated way, also taking into account other parameters. On the other hand, we argue that even if proposals are on a par in terms of those other parameters, there are other, more appropriate ways to sort proposals in light of the approvals they have received.
\end{abstract}

\keywords{digital democracy, online deliberation, participatory democracy, deliberative democracy, sorting, digital democratic innovations}

\section{Introduction} \label{sec:introduction}

Recent attempts to improve democratic participation often make use of the internet \citep{mossberger2008digital,Landemore-OpenDemocracy}. In particular, in the last decade numerous online deliberation platforms (DPs) have been developed and put into practice. These platforms let citizens engage in all sorts of processes including citizens' assemblies \citep{Parsons2019DigitalAssemblies}, participatory budgeting \citep{Matheus2010CaseEngagement, Holston2016EngineeringCase, Mre2021IncreasingProcesses}, drafting policy \citep{Randma-Liiv2022EngagingEurope} and determining the public opinion \citep{Small2021Polis:Spaces}. DPs thus represent an instance of democratic innovations: institutions designed to increase democratic participation \citep{Smith2009DemocraticParticipation, elstub2019defining, Mikhaylovskaya2024EnhancingInnovations}.

A successful example of one such platform is \platform{Consul}, which the Madrid city council has been using since 2015. This platform enables citizens to submit proposals for improving city life. Proposals that get the support of 1\% of the citizens go to a binding public poll. 
Between 2015 and 2019, a total of 27,309 proposals were made on the platform \citep[p.~160]{Pina2022DecideE-participation}. Similarly, the \platform{Decidim} platform, developed and maintained in Barcelona, has been used to support online participatory processes across the world, with over 3,5 million registered users in Brazil \citep{BarandiaranDecidimScience}.

As these examples illustrate, the advantage of increased participation on DPs comes with a major challenge: managing the stream of information that they generate. This necessitates several design choices that have significant impact on the democratic process.\footnote{Barandiaran et al., developers of the \platform{Decidim} platform, emphasize that decisions regarding software design, especially proposal sorting, carry political implications \citeyearpar[p.~68]{BarandiaranDecidimScience}. Similarly, Mikhaylovskaya \citeyearpar[p.16]{Mikhaylovskaya2024EnhancingInnovations} argues that well-designed algorithms on DPs can enhance inclusiveness and representation for marginalised groups. However, there has been limited discussion about the democratic values that should guide the design of DPs \citep{Smith2019ReflectionsInnovations, Mikhaylovskaya2024EnhancingInnovations}.} In particular, and in line with the \emph{deliberative turn} in political theory \citep{Dryzek2002deliberativeturn, palumbo2024deliberative}, most of these platforms have a module that accommodates a \emph{deliberation phase} where citizens can post their own proposals for a given topic or area, view and comment on proposals of others, and evaluate proposals by attaching ``likes'' to them. For the most successful DPs, the number of proposals for a given query is often in the order of magnitude of 100s, 1000s or even 10.000s (see also Section \ref{sec: Sorting in practice}). Arguably, in such cases, \emph{if and where} a proposal ends up on the next user's screen has an impact on the attention it gets and thereby influences the deliberation and thus the decision-making process as a whole.

This raises an important normative question: in what order should we show proposals, if the use and output of these platforms is supposed to be \emph{democratic}?\footnote{Our research question closely aligns with the field of recommender systems (e.g., \citep{jannach2010recommender, aggarwal2016recommender}), which address information overload by providing personalised recommendations for i.a.\ products, news or friends. However, recommender systems aim to give recommendations that are specifically linked to an individual user's past clicks and views, whereas in this paper we focus on how to show information to an arbitrary ``next user'' or to a group of such users. Moreover, recommender systems usually serve commercial rather than democratic purposes, even though recently researchers have started to pay attention to matters of fairness and diversity  \citep{Zhao2023FairnessSurvey, Vigano2023TheSystems}.} This question has received limited attention both in practice and in the academic literature. Existing platforms often sort proposals in an \emph{ad hoc} manner and do not explicitly discuss how and why they choose a certain method. Moreover, most platforms do not distinguish between various purposes the list of proposals may serve and tend to focus on a single parameter when sorting proposals. Political philosophy and philosophy of technology has, to the best of our knowledge, also remained silent on this question. 

In this paper, we make a start with the normative evaluation of sorting methods for online deliberation. First, we highlight the importance of a principled approach to sorting methods (Section \ref{sec: sorting methods: A Principled Approach}). In preparation for this, we introduce a conceptual framework that allows us to distinguish sorting methods in terms of their purposes and the parameters they take into account. Second, we use this framework to study a representative sample of successful DPs in terms of their sorting methods (Section \ref{sec: Sorting in practice}). We observe that in practice there is great variety in the way different platforms sort proposals, while there is usually no explicit justification for the chosen sorting methods. 
Third, we advance two specific arguments and thereby shed critical light on the common approach of sorting proposals in terms of their number of approvals (Section \ref{sec:popularity}). Our arguments draw on general principles and values of deliberative democracy as discussed in political theory. On the one hand, we show that \emph{if} approvals are to be taken into account, this should be done in an integrated way, also paying attention to other parameters related to metadata and content. On the other hand, we argue that even if proposals are on a par in terms of those other parameters, there are other, better suited ways to sort proposals in light of the approvals they received. 
We conclude with a survey of questions for future research on the design and implementation of sorting methods on DPs.

\section{Sorting Methods: a Conceptual Framework} \label{sec: sorting methods: A Principled Approach}

In this section, we first explain what we mean by sorting methods and specify various aims of these methods. We argue that at least for the purpose of online deliberation, the choice of sorting methods is important and non-trivial. Next, we discuss the parameters that can inform a given sorting method.

 \subsection{Sorting Methods} \label{sec:Ranking Rules and Methods}

\begin{figure}[htbp]
    \centering
    \includegraphics[width=\textwidth]{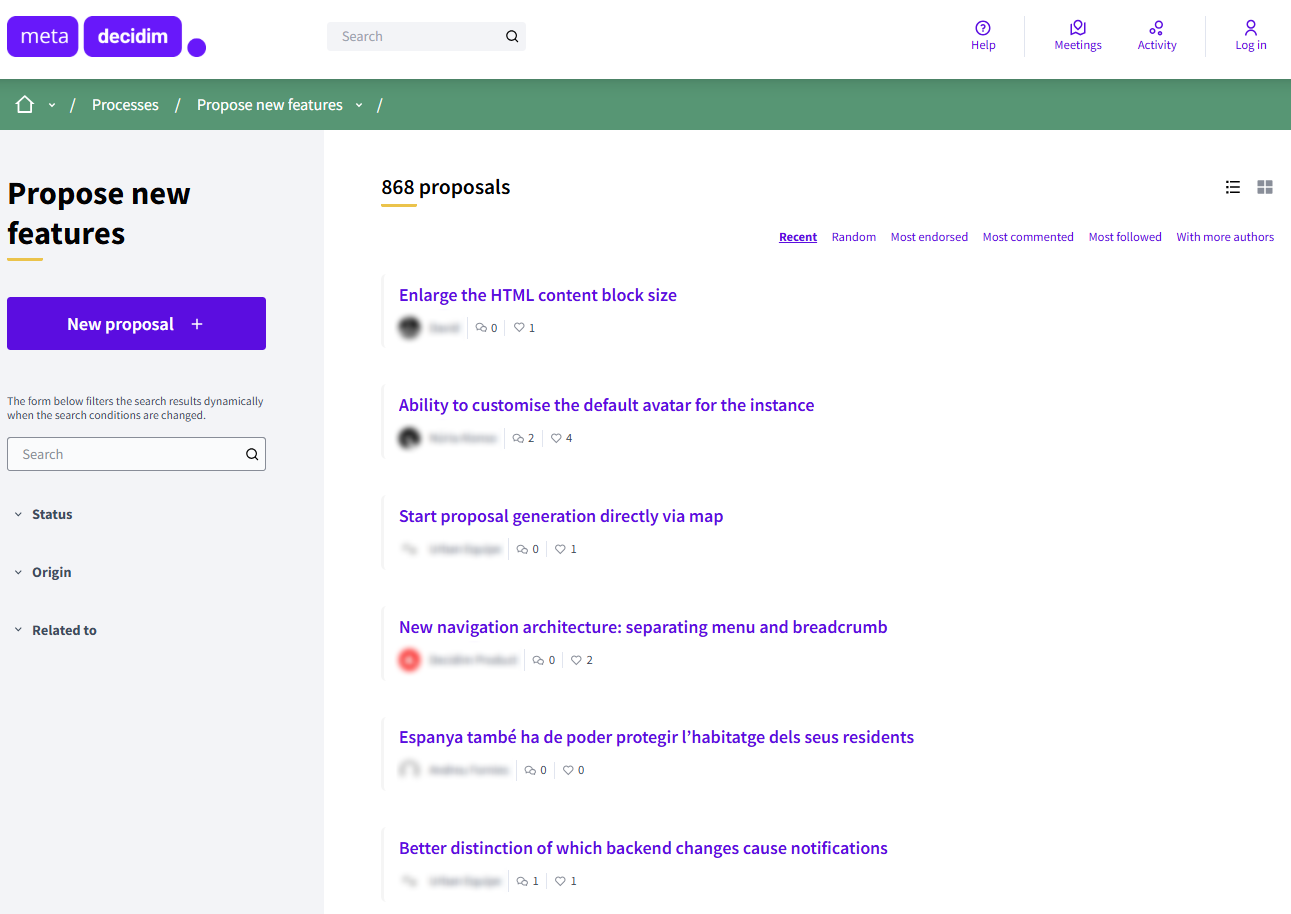}
    \caption{Project overview page with a list of 868 proposals. Screenshot taken on meta.decidim.org on August 5 2025 \citep{decidimproposals}.}
    \label{fig:decidim-proposals}
\end{figure}

\begin{figure}[htbp]
    \centering
    \includegraphics[width=\textwidth]{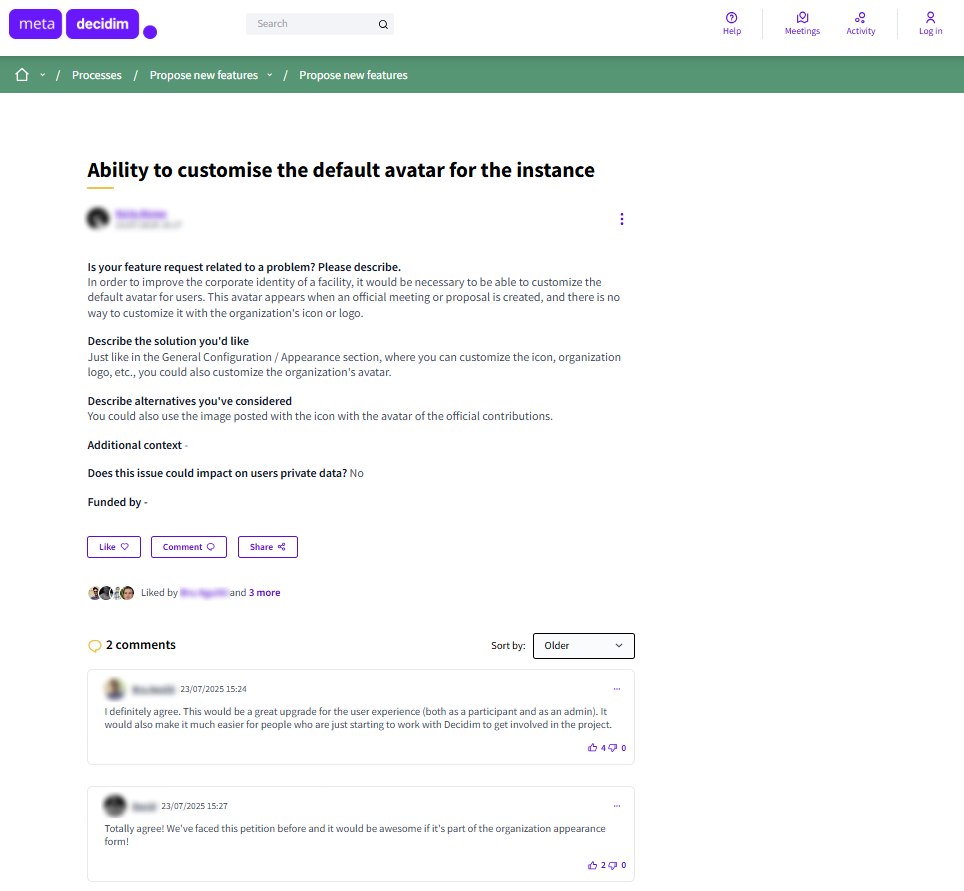}
    \caption{Detailed view of a single proposal on \platform{Decidim}, including its description, number of endorsements, and comments. Screenshot taken on meta.decidim.org on August 5 2025 \citep{decidimproposals}.}
    \label{fig:decidim-proposalexample}
\end{figure}

Recall that a distinctive feature of DPs is that they allow citizens not just to vote on pre-given policy proposals or laws, but also to come up with new proposals themselves. 
In typical instances of DPs, these proposals are meant to address a fairly general question such as ``How would you improve your neighbourhood?'' for a participatory budgeting project, or ``How can universities contribute to the mental health of students?'' for a university-wide poll. Proposals for the first of these two examples could be titled ``Organise a weekly language support class for citizens'' or ``Invest in more benches in the park''. Typically, users can post a short description of their main idea, indicating how the proposal should be carried out, what they expect it to cost, why they think it is important or useful, etc. Figure \ref{fig:decidim-proposals} shows a list of such proposals on \emph{Metadecidim}, a website for the community of \platform{Decidim} users where they can propose specific improvements to the software itself. Figure \ref{fig:decidim-proposalexample} provides an illustration of a selected proposal, as well as some of the comments it received.

Our starting point is that the proposals that have been posted up to a certain point in time are to be shown to users in the form of a list. This list typically ranges over several pages in any standard browser format.\footnote{\label{footnote:otherformats} Showing users a complete list of all proposals is most common in practice, but other representation methods are also exist. For example, Polis (\url{https://pol.is/home}) and Make.org (\url{https://make.org}) show users a small selection of proposals that differs per user. Another way is to show proposals on a map, as is often done in participatory budgeting projects. Finally, proposals can be shown in grid or tabular interfaces\citep[p.~1498]{Lewandowski2021FactorsResearch}. While many of our findings apply \emph{mutatis mutandis} to these alternative formats, we leave their exploration for future work.} For instance, on \platform{Metadecidim}, and at the time of writing, 868 proposals have been posted in total. By clicking on items in this list, users can read more about them, comment on them, and evaluate them e.g.\ by posting ``likes''. In what follows, we use the term ``approvals'' to refer to the (positive) expressions of support by users on proposals.

We define a \textit{sorting method} very broadly, as any algorithm that generates such a list of proposals on the basis of the information that is available at the platform, at any given point in time. One may e.g.\ sort the proposals by their number of approvals, by their time of posting, or in a purely random way.

\subsection{Purposes of Sorting} \label{sec: sorting goals}

Which sorting method is most suitable for a given DP depends \emph{inter alia} on the overall purpose the list of proposals is supposed to serve. Based on our survey of a representative sample of successful DPs --- see Section \ref{sec: Sorting in practice} ---, we distinguish three different, broad purposes that may inform the choice for a sorting method. 
First, consider a process in which users are invited to continuously upload, approve, comment on, and revise or amend proposals.\footnote{On some platforms, such as \platform{Decidim} and \platform{Parta}, uploading and approving proposals is kept separate. That is, in a first phase, users can only upload proposals, and in a later stage they can approve and comment on proposals. On other platforms, such as \platform{Go Vocal}, \platform{Your Priorities} and \platform{LiquidFeedback}, submitting new proposals and approving them happens simultaneously. \label{footnote:seperate}} This process is typically referred to as the ``deliberative phase'' of a project's decision-making cycle, and is seen as a main point of entry for large-scale, online participation. Throughout this deliberative phase, we need to show each user a list of those proposals that have already been posted and are in that sense ``up for debate''. We call this \textit{sorting for online deliberation}.

After the deliberation phase, there are several possible next stages in decision-making, some of which require a sorting method too. One option is to use the results of the deliberation phase to inform an in-person committee meeting. In this case, a sorting method may be used to inform the committee which proposals it should discuss. We call this \textit{sorting for offline deliberation}. Alternatively, the results of the online deliberation phase may be taken to a vote.\footnote{A decision process might go through a few cycles of online deliberation, offline deliberation, or voting. For example, at a participatory budgeting process on \platform{Decidim}, after a first round of proposals and voting on them, a selection of proposals may be evaluated and amended by a group of public servants in terms of their feasibility and cost, after which another round of online commenting and voting is enabled [personal communication, Andrés Pereira de Lucena, July 2 2024]. } Here, \textit{sorting for voting} is meant to settle the order of the options over which the votes are cast.\footnote{Note that this says nothing about the type of ballots in voting. The ballot itself may also take the form of a list, but it may also be a single option or a set of options.} In both cases, sorting typically goes together with some form of \emph{shortlisting}, e.g.\ by imposing some upper bound on the number of eligible proposals.

Let us illustrate the above. One can see the submission and evaluation of proposals on  \platform{Consul} and \platform{Decidim} as a form of deliberation. During this phase, no decisions are being made, but instead citizens are encouraged to reflect upon which problems live in the city, to view arguments for or against certain proposals, and to post new ones. Sorting proposals in this phase is thus an example of sorting for online deliberation.

Next to the possibility to freely submit proposals to improve the city, both platforms also support a module for participatory budgeting, enabling citizens to decide how to spend a given budget through various projects. First, citizens can upload proposals, approve and comment on them. This is another instance where sorting for online deliberation takes place. In the second phase, users can vote to decide which proposals should be funded. Sorting proposals in this phase is an example of sorting for voting.

An example of sorting for offline deliberation can be seen on the implementation of the \platform{OpenStad} platform in one specific neighbourhood in The Hague, where a ``neighbourhood agenda'' is created on the basis of posts that have been submitted and evaluated by regular citizens.\footnote{See \url{https://openstad.org/voorbeelden/den-haag-wijkagendas}.} Here, the idea is that citizens themselves should determine which issues and proposals should get priority at the meeting of the neighbourhood council. On the basis of the ``likes'' that have been submitted, the platform then determines which items go on to the council.

Sorting methods can thus serve at least three different purposes. For each of these, one may in principle use a different sorting method. Arguably, the choice for a given sorting method \emph{depends} for an important part on this purpose. In particular, when sorting for offline deliberation in some plenary meeting, one typically aims for a unique shortlist that serves as the agenda for the physical meeting in question. This agenda should arguably reflect the size of groups that support certain proposals, and thus be ``proportional''.\footnote{We discuss proportional sorting methods in more detail in Section \ref{sec:popularity}.} The latter requirement also applies to sorting for voting, even though there one can in principle randomize the sorting of selected proposals across different users in order to rule out certain biases. When sorting for online deliberation, such diversification is also possible. Moreover, as explained above, the sorting for online deliberation will typically change over time as new proposals are posted or users evaluate existing ones. Finally, as we will argue in Section \ref{sec:popularity}, proportionality is less of an issue for (online) deliberation, where it is more important to ensure that all participants are ``being heard'' at least to some extent.

In the remainder of this paper, we focus on sorting for online deliberation. That is, our main question in this paper is how proposals should be sorted during online deliberation, and what normative principles  can ground a sorting method for this specific purpose. The reasons for this focus are threefold. First, deliberation is the starting point of many decision-making processes facilitated by DPs. As such, it plays an important role in determining the further course of these processes: it sets the agenda and determines what the options are for voting or further discussions. Second, sorting for online deliberation and sorting for offline deliberation are much less explored than sorting for voting.\footnote{Sorting for voting has been studied in the context of traditional (offline) elections, where a \emph{ballot order effect} has been observed: often, candidates gain a signficiant extra percentage of votes by being sorted among the first candidates. This effect is not universal and depends on the information voters have about candidates and whether those candidates belong to major or minor parties \citep{MillerKrosnick-1998-ballotordereffects,Ho-Imai-2008,Pasek-etal-2014}.} Third, the sorting method for offline deliberation tends to be relatively opaque and hard to trace, and typically depends on the city or institution hosting the deliberation, making it hard to obtain any general insights at platform level. 

In the specific case of sorting for online deliberation, it should be highlighted that the sorting method is applied repeatedly while the deliberation process continues. After all, proposals can be submitted and approved continuously.\footnote{See however footnote \ref{footnote:seperate} for some exceptions.} This means that in principle, every time a new proposal or a new approval is submitted, a new list could be created. In practice, the list of proposals is updated on a daily or weekly basis. In any case, the application of the sorting method is not a single instance, but it is an operation that happens over and over again throughout the deliberation phase.

\subsection{The Need for a Principled Approach} \label{sec:The Need for a Principled Approach}

One might wonder why we should care about which sorting method is being used, or why the answer is anything but obvious. Why can we not just sort the proposals by their date of submission, by their popularity, or simply randomly?

Although no systematic experimental work has been carried out on the design of sorting methods, we have good reasons to think that they do have an impact on the deliberation process and its result. First, recall that evaluating and commenting on proposals is an indispensable part of online deliberation. However, given that users have limited time and mental capacity, they cannot evaluate \textit{all} proposals. As we show in Section \ref{sec: Sorting in practice}, typical instances of online deliberation feature several hundreds or even thousands of proposals. Studies on search engine behaviour show that, when users are subjected to such long lists, most attention goes to results on top of the page \citep{Lewandowski2021FactorsResearch}. Furthermore, a majority of individuals look no further than the first results page \citep[p.~121]{Spink2004WebWeb}. Therefore, we have good reasons to assume that the sorting method influences how well each proposal is inspected.

This means that if we sort proposals based on submission date, the success of a proposal will depend on when it was posted. Old proposals will either be ``top of the bill'' throughout, or get buried under new proposals. This strongly invites strategic manoeuvring, such as trying to submit a proposal right after the call opens, re-submitting (clones of) a proposal every other day, or waiting until right before the call closes to have one's proposal in a most visible position. Such strategic manoeuvring goes against the very idea that, by making deliberation asynchronous, DPs can contribute to the fair and equal participation of large numbers of citizens in the decision-making process. If timing really matters, then it no longer holds that anyone can contribute to the discussion whenever it suits them best.

If instead we sort proposals based on their number of approvals, then large, homogeneous groups of users may get an undue advantage. After all, a single group may approve a large number of proposals, and thereby bury all other proposals approved by smaller groups, making sure those end up very low in any of the generated lists. Moreover, since new proposals can be introduced all along, it is very hard to compare the scores of different proposals unless we also take into account how often they have been inspected. We explain and discuss these problems in detail in Section \ref{sec:popularity}, where we also explain how they can be avoided in a principled way.

The above considerations do not apply if we would generate random lists per user. However, while it may perform better than sorting by popularity or by date, random sorting does so in a brute force and unsystematic way. Random sorting moreover has problems of its own. For individual users, random sorting means that whatever proposals they see -- say, in the top twenty of their list -- may be approved by hardly anyone, and so these proposals may not be very qualitative or representative of the broader society. Conversely, it also means that we lose out on an obvious way in which users can have an impact during the deliberation phase, viz.\ by ``upvoting'' some proposal so that it gets more attention. Both these elements may negatively affect the motivation of individual participants to use the platform. Moreover, if the ratio of the number of users to the number of proposals is not large enough, then although all proposals are ``treated equally'' \emph{ex ante}, \emph{ex post} some proposals might have a higher visibility due to a purely random factor.

A key motivation for moving deliberation online is to increase the quality and (perceived) legitimacy of democratic decisions \citep{deZeeuw2020AParticipation, Rumbul2019TheEngagement, mossberger2008digital, Escher2023EffectsMunicipalities}. As we have argued, the sorting method that is used in online deliberation has a significant impact on which proposals receive high visibility. Employing a specific sorting method may give an undue advantage to already influential groups in the decision-making process, or allow arbitrary factors such as the time of posting to impact outcomes. If this pattern persists, it could demotivate certain groups from participating, undermining the democratic ambitions of DPs. Therefore, there is a need for a principled approach to sorting methods that aligns with the core objectives of DPs and deliberative democracy more generally.

\subsection{Sorting Parameters}\label{subsec:fourpara}

Let us now move away from the simple methods discussed in the previous section. On what basis could proposals be sorted during online deliberation? It will be useful to distinguish three broad categories of parameters that may inform the sorting of proposals:
\begin{itemize}
\item[1.] The \emph{content} of proposals;
\item[2.] The \emph{evaluation} of proposals (by previous users);
\item[3.] All \emph{metadata} of proposals on the platform'.
\end{itemize}

We discuss these one by one. First, \emph{content} refers to what is actually being proposed. Minimally, a proposal will consist of a title and some description. However, in most instances of DPs, richer formats are available: one may for instance add certain labels or tags to a proposal that refer to a general theme or political issue they try to solve, or to the specific region, neighbourhood, or social group that would benefit from the proposal at hand. Such labels may either be user-generated or introduced by the platform hosts. 
Finally, in case proposals concern concrete projects (e.g. in city development) one may include an estimation of their cost.

Content-based sorting could e.g.\ aim for diversity in terms of the proposals.\footnote{In the absence of specific classifiers or information about the cost of proposals, this could involve some form of interpreting the description of the proposal, whether by humans or through an AI.} This would allow us to avoid that, say, the first fifty proposals in the list are all concerned with a single neighbourhood in the city, or all of them only address one very specific sub-problem of the general issue under discussion. A lot is possible here, but so far this area seems mostly unexplored both theoretically and in practice --- safe for the cost parameter that we mentioned in the previous paragraph.\footnote{As we explained above, some platforms just show the proposals on a map instead of listing them. While this has the advantage that geographic information is made clear, it has the disadvantage that many users may be inclined to only focus on their own neighbourhoods or regions.}

By the \emph{evaluation} of a proposal we mean expressions of preference, support, or (dis)approval by users concerning proposals. Most often, DPs take inspiration from familiar social media and allow that proposals can be ``liked'' by users. One platform, i.e.\ \platform{LiquidFeedback}, goes one step further and allows users to submit ``conditional'' approvals, expressing under which conditions they would be willing to support a given proposal. 
One can of course also imagine other, richer formats where e.g.\ proposals are graded or even rated on different criteria, though at present we know of no DPs that do this. What matters ultimately is that such evaluations by past users may be taken into account when showing them to the next user. 

Finally, by \emph{metadata} we refer to any other information about the proposals that is not part of their content, but rather concerns their history on the very platform: when and by whom they were posted, how often and by whom they have been viewed, etc. One common way to sort proposals is simply by their date of submission (either starting with oldest or with newest), but some platforms also prioritize proposals that have received more comments in some determinate (recent) span of time. 

It should be noted that, depending on how we exactly interpret and use them, \emph{comments} can be seen as giving rise to information of all three types. While the number and origin of comments constitutes a particular set of metadata, comments often also carry evaluative information and content in and by themselves. Since the platforms that use comments for sorting rely on the \textit{number} of comments, we have classified this method under \textit{metadata} in our survey (cf.\ Table \ref{tab:overview} in Section \ref{sec: Sorting in practice}). 

In our study of existing DPs, we will consider each of the above-mentioned parameters and explain how they are implemented currently. In addition, we will also register at what exact level the sorting method is chosen. First, the \textit{developers} of the platform may simply build in one sorting method, or give the choice among several methods to those using the platform. Among the latter, we may further distinguish between the platform's \textit{hosts} -- typically, some company, municipality, or political party that uses the platform for their organisation -- and its \textit{users} -- the ``demos'' that is supposed to provide the input in the form of proposals, approvals, comments, and votes. While we do not return to this choice in our normative evaluation, it is an open question at what level the sorting method \emph{should} be chosen. In any case, if we do offer this choice to platform users or hosts, then it is important that they are fully informed of what the normative presuppositions and practical implications of any given method are.

\section{Sorting Methods in Practice} \label{sec: Sorting in practice}

Let us now consider how DPs in practice sort proposals for online deliberation. Drawing from two online databases \citep{DemocracyFoundationListProjects, DemocracyTechnologiesCollectionParties}, we selected eight DPs based on the following criteria:
\begin{itemize}
\item The platform allows users to propose proposals, (dis)like proposals, and show proposals to the users in a list-like format.\footnote{See footnote \ref{footnote:otherformats} for a brief discussion of other formats.}
\item The platform has been implemented for governmental, municipal or political deliberation and decision-making.
\end{itemize}
So while ours is not an exhaustive overview of all the available DPs, it does form a representative sample of successful and widely used platforms, with a specific focus on sorting methods. 

\paragraph{Numbers that Count} The number of proposals per use case on these DPs is typically so large that some form of sorting method becomes essential. To illustrate this claim, we made a selection of 19 cases across four platforms (\platform{Decidim}, \platform{Consul}, \platform{Go Vocal}, and \platform{Your Priorities}) in various cities and countries, as described on the platforms' websites. The complete list of cases including the sources is depicted in Table \ref{tab:proposal-numbers} in Appendix \ref{appsec:numbers}.  
We found an average of 2724 proposals per case, with a median of 628. The number of proposals ranged from 83 to  18859. These data highlight a consistent pattern: regardless of the specific use case, the number of proposals tends to exceed what users can realistically evaluate. Even ‘as few as' 83 proposals would likely be too many for a typical user to read.

\paragraph{Sorting Methods in Practice} Let us now consider how, on these DPs, proposals are being sorted. First, it is notable that a majority of the platforms do not use a different sorting method depending on the purpose of the list. Compared with sorting for deliberation, only 2 out of 8 platforms used a different method for voting. \platform{OpenStad} makes a distinction between sorting for online deliberation and for voting [I. Rieken, personal communication, June 10 2024]. During the online deliberation phase in participatory budgeting projects, it is common to let users choose to order proposals based on date, approvals, comments or cost. By default, the order is set on ‘newest first’. In contrast, during sorting for voting, proposals are shown in a random order. Also \platform{Consul} offers different sorting options during participatory budgeting, depending on the phase. During the deliberation phase, users can order proposals randomly or based on most approvals. During the voting phase, users can order randomly or by cost [S. Strohmenger, personal communication, June 21 2024]. In a similar vein, \platform{Decidim}'s default sorting method varies depending on the stage of the process. During the proposal evaluation (i.e. online deliberation), the default method is random sorting, whereas, after evaluation, the default shifts to sorting based on the number of approvals. Additionally, \platform{Decidim} offers different sorting options depending on the participatory process. 
In the deliberation phase for \textit{proposals}, users have the options shown in Table \ref{tab:overview}, without the cost option, and including a ‘most followed' and ‘most authors' option. Meanwhile in the deliberation phase of \textit{participatory budgeting}, users can only choose between random, highest or lowest cost [A. Pereira de Lucena, personal communication, June 26, July 2 \& 8 2024]. 

As concerns sorting for online deliberation, our findings are summarised in Table 3 
in Appendix \ref{appsec:sortingmethods}.\footnote{Our observations are based on the following sources.
\platform{Adhocracy+}: Personal communication with Tietje Khieu [June 24 2024], Janek Gulbis [July 29 2025] and online demo \citep{adhocracy}. \platform{Go Vocal}: online guide \citep{govocal}. \platform{Consul}: online guide \citep{consuldemo,consulguide} and Personal communication with Simon Strohmenger [June 21 2024]. \platform{Decidim}: online demo and guide\citep{decidemdemo,decidemguidecomments}, and Personal communication with Andrés Pereira de Lucena [June 26 2024, July 2 \& 8 2024]. \platform{Your Priorities}: online demo \citep{yourpriorities}. \platform{LiquidFeedback}: book \citep{Behrens2014TheLiquidFeedbackb}. \platform{OpenStad}: Personal communication with Ian Rieken [June 10 2024]. \platform{Parta}: Personal communication with Mathijs Kemp [June 7 2024].} Three general facts stand out. First and foremost, there is no universally preferred (default) sorting method across the platforms. Each DP appears to favour a distinct approach, including sorting by date (\platform{Adhocracy+}, \platform{Your Priorities}), randomly (\platform{Decidim}, \platform{Parta}), proportionally (\platform{LiquidFeedback}), or using a combination of multiple parameters (\platform{Consul} --- this is referred to as `combi' in Table \ref{tab:overview}).

Second, a majority of the DPs (6 out of 8) enable users to change the default sorting method. While on these platforms it is possible for the host (e.g. a municipality) to disable this function for users, which would entail that the hosts decide, we see that most commonly they leave this choice for users. So most DPs are designed with the (implicit) assumption that we can or should let the users choose on what basis proposals should be sorted. Only two DPs (\platform{Parta}, \platform{LiquidFeedback}) have just one available sorting option, and thus the developers (Dev.) decide.

Third, the most frequently offered methods that users can choose from are sorting on date (newest first, 6 platforms), on approval numbers (most approvals first, 6 platforms)\footnote{In case the platform also allows disapprovals, what matters is the approval/disapproval ratio. This is for example the case for the debates on \platform{Consul} [S. Strohmenger, personal communication, June 21 2024].} and randomly (5 platforms). Note that each of these methods only take into account (at most) one specific parameter, rather than merging different types of information available on the platform.

There are two exceptions to this general observation. In particular, \platform{Go Vocal} claims to offer a sorting option that takes into account ``approvals, activity, date and engagement'' \citep{govocal}.\footnote{We have not been able to trace back exactly what these terms refer to, despite repeated inquiries to \platform{Go Vocal}'s organisation.} \platform{Consul} offers an option called ‘most active’, which shows the proposals that have been supported the most in the last few days [S. Strohmenger, personal communication, June 20 2024]. It thus takes into account both a time-frame, which can be configured by the administrators, and approval numbers. On various platforms (e.g. \platform{Consul}, \platform{Adhocracy+}) it is also possible to search actively for particular proposals, based on search words or available tags.

Beyond these general observations, there is a lot of variety among platforms. Again, three observations stand out. First, in addition to the sorting methods that were mentioned in the previous paragraph, various platforms offer more rare sorting options, such as oldest first, least number of approvals first, most comments first, based on number of authors or followers (\platform{Decidim}), and – for participatory budgeting – based on cost (highest or lowest first).  

Second, while randomising is a common sorting option, platforms differ in how this is implemented exactly. For instance, \platform{Go Vocal} creates a random ordering of proposals, which is the same for all users and changes once a day \citep{govocal}. For \platform{Parta}, a random order means that a user will see a different order each time the list is viewed [M. Kemp, personal communication, June 7 2024]. On \platform{Consul} and \platform{Decidim}, the random order is saved in the browser.\footnote{\platform{Consul} offers randomisation only for participatory budgeting projects, both in the deliberation phase and in the voting phase [S. Strohmenger, personal communication, June 20 2024].} Thus, for the same person using the same browser, the order will be consistent until the session expires -- which happens when someone e.g. closes the browser or logs out of their account.

Third, one platform stands out for the way it handles approvals in sorting proposals: \platform{LiquidFeedback}. For \platform{LiquidFeedback}, all proposals are sorted using the so-called \textit{Harmonic Weighting} algorithm \citep[p.~74]{Behrens2014TheLiquidFeedbackb}. This sorting method aims to create a list that reflects users' evaluations, as indicated by their approvals, proportionally (cf.\ Section \ref{subsec:twoalt}). Earlier versions of \platform{LiquidFeedback} sorted proposals based on approval numbers, combined with tie-breaking in favour of the oldest proposals \citep{Behrens2014TheLiquidFeedback}. However, due to shortcomings that the designers of \platform{LiquidFeedback} experienced with this method, the developers of this platform shifted to Harmonic Weighting.

\section{Most Approvals First? A Critical Evaluation}\label{sec:popularity}

Recall that on most platforms, a common option is to sort proposals by their number of approvals. This may well be the most intuitive approach to many, and it certainly is most familiar from social media platforms. However, in what follows, we show that this method is problematic in two ways, in light of basic principles of deliberative democracy (Section \ref{subsec:consid}). On the one hand, we argue that approvals should be combined with information about the content and metadata of proposals (Section \ref{subsec:integrating}). On the other hand, we argue that even if proposals are on a par in those other respects, sorting them by their number of approvals is not the most appropriate way of handling approvals (Section \ref{subsec:twoalt}).

\subsection{Normative Considerations for Online Deliberation}\label{subsec:consid}

What properties should (online) deliberation have, in order to be democratic? To answer this question, we need to say what it is about deliberation that makes it valuable in the first place. On the one hand, \emph{procedural} arguments for deliberative democracy invoke properties such as inclusion, reflection, and dialogue as essential for the legitimacy of political decisions and processes leading up to them. On the other hand, deliberative democracy has been advocated using \emph{instrumental} arguments regarding the quality and perceived legitimacy of decisions. Let us go over these considerations one by one.

Following Pettit \citeyearpar{Pettit-2001}, we distinguish three core properties that are taken to be necessary (if not sufficient) for a procedure to render legitimate decisions, according to the deliberative democracy paradigm: democratic procedures should be (i) \emph{inclusive}, i.e.\ involve the active participation of all those affected by the decision in question; (ii) \emph{judgmental}, based on arguments and supported by a critical, open-minded attitude towards one’s own views and preferences; and (iii) \emph{dialogical}, putting talking rather than voting at the centre, with the aim of achieving mutual understanding if not some form of (meta)agreement.\footnote{Many similar characterisations can be found in the literature on deliberative democracy. In their introduction to the Oxford Handbook of Deliberative Democracy, B\"achtiger and co-autors write: ``[W]e define deliberation itself minimally to mean mutual communication that involves weighing and reflecting on preferences, values, and interests regarding matters of common concern'' \citep[p. 1]{Baechtiger-etal-2018}. In their chapter on Online Deliberation, Strandberg and Gr\"onlund write: ``[T]he general consensus appears to be that the communication should be conceptually linkable to deliberative theory and four core pillars: that discussions must be inclusive, rational-critical, reciprocal, and respectful'' \citep[p.\ 366]{StrandbergGroenlund-2018-onlinedeliberation}.}

Deliberative democrats have long stressed the importance of having a dialogical exchange ``of all, with all’’, before reaching a decision. This means both that everyone should have a fair chance of raising their concerns, and that everyone should listen respectfully to whoever gets to speak. In particular, \emph{size should not matter}: minorities should be heard just as well as majorities.\footnote{In his seminal paper, Manin \citeyearpar[p.~352]{manin1987legitimacy} writes: ``As political decisions are characteristically imposed on \textit{all}, it seems reasonable to seek, as an essential condition for legitimacy, the deliberation of \textit{all} [...]''. Similarly, Chamberlin and Courant state that ``representativeness in deliberations is measured by the degree to which one's interests will be faithfully entered into the deliberations, and not by the frequency with which this may occur nor by the likelihood that they will prevail in the subsequent social decision'' \citep[p.~723]{Chamberlin1983RepresentativeRule} and ``To deny minority viewpoints an airing in the deliberations unnecessarily is to undercut the legitimacy of the ultimate decisions'' \citep[p.~719]{Chamberlin1983RepresentativeRule}. More recently and in the specific context of digital democratic innovations, Mikhaylovskaya \citeyearpar[pp.13-14]{Mikhaylovskaya2024EnhancingInnovations} emphasises that ``marginalised and vulnerable groups should be included in such a way as to not be disadvantaged by other interest groups.''}

The judgmental aspect similarly implies that for proper deliberation, the involved citizens should be exposed to \emph{all} (sensible, relevant) viewpoints, perspectives, and considerations whenever possible. For the specific case of sorting or selecting proposals, we thus conclude that diversity is key: rather than trying to find the best or most popular proposals, it is important that during deliberation, users are exposed to diverse proposals. 

Such diversity is also crucial for the dialogical aspect of deliberative democracy. It has long been acknowledged that in practice deliberation may be indecisive and typically has to be followed by a voting round or some other form of aggregation. Nevertheless, various authors have argued that deliberation does aim to create mutual understanding and meta-agreement \citep{NiemeyerDryzek-2007}: agreement on how the problem is to be conceived, what feasible options are, what values are relevant (even if we may disagree on their priority order), what factual statements are relevant for the decision at hand, or what is an appropriate way to settle the decision in the absence of full-blown consensus. So the dialogical aspect of deliberative democracy requires not just dialogues among like-minded citizens, but instead the confrontation of different viewpoints and perspectives in order to end up with a commonly acceptable problem definition.

As concerns instrumental arguments for deliberation, we focus on two in particular: (i) on the \emph{epistemic} account of democracy, deliberation enhances the quality of decisions \citep{GoodinSpiekermann-epistemictheoryofdemocracy-2018,Landemore2012DemocraticReason,DietrichSpiekermann-2013}; and (ii) deliberation fosters \emph{perceived legitimacy}, i.e.\ it leads to decisions that are \emph{perceived} as legitimate by those affected by the decisions, and citizens may therefore be more willing to accept them \citep{Escher2023EffectsMunicipalities}.\footnote{In the literature, this concept is known by various other terms: \textit{descriptive legitimacy}, \textit{output legitimacy} \citep{10.1145/3665332}, \textit{legitimacy beliefs}  \citep{Escher2023EffectsMunicipalities}. In turn, these notions are tightly linked to the notion of trust (in decisions, in government, or in the political system as a whole) \citep{mossberger2008digital, terwel2010voice}.} In as far as either (i) or (ii) is seen as what makes deliberation valuable, we may well want to ensure that online deliberation has similar effects.\footnote{More recently, an argument has been made that interactions via online deliberation can be structured and implemented so as to foster specifically democratic virtues \cite{Anna&Elise-nurturing}. While we are sympathetic to this viewpoint, the discussion in \cite{Anna&Elise-nurturing} focuses mostly on advanced, often AI-supported functionalities for this purpose: auto-translation, moderation, prompting users to take into account certain (counter)arguments, etc. It remains to be seen whether the choice of sorting methods can also be motivated in terms of democratic virtues.}

When considering the deliberation phase on a DP, epistemic arguments for deliberation pertain to the quality of proposals that are being produced: whether they are realistic, balanced, compatible with established facts, etc. If the choice of a sorting method has a positive impact on this, this obviously counts in its favour. Likewise, if a sorting method is such that it is perceived as legitimate and appropriate for the setting of online deliberation, then this is a pro for that method.

\subsection{Integrating Approvals and Other Parameters}\label{subsec:integrating}

We now argue that \emph{if} approvals are to be taken into account, then this should be done in an integrated way: evaluative information must be interpreted in light of the content of proposals and their metadata.

Consider first the metadata about who posted or actively contributed to a given proposal. Suppose that a large number of proposals by one author get most approvals. This may be evidence for the quality of those proposals, but it may also say something about the rhetorical or digital skills of this one author, or about their popularity among users. In this case, sorting by approval numbers goes against the inclusiveness requirement (cf.\ Section \ref{subsec:consid}). 

Similarly, the extent to which a proposal has been viewed is crucial to determine the significance of its evaluations. Recall that on the epistemic argument for deliberative democracy, deliberation has a positive influence on the quality of proposals. Approvals may well be taken as a proxy for such quality, but only when the proposals have been inspected by a sufficient number of users. So if we decide to give less exposure to proposals in light of their number of approvals, we should only do so once it is clear that they have been well-inspected. Obviously, it remains to be seen how exactly we should measure the degree to which a proposal was inspected. However, unless we take this specific metadata into account, we run the risk of creating a feedback loop in which mediocre proposals keep on being sorted above better ones.

Turning to the content of the proposals, when the most popular proposals are tied to certain regions, neighbourhoods, or specific labels, it would be strange to just ignore this information and stick to listing the most approved ones first. Instead, we may first diversify across such content-related aspects of the proposals, and only then take into account approval numbers. 

The importance of the above considerations becomes even more salient when considering an important and well-established empirical fact, which is typically referred to as the \emph{digital divide}.\footnote{Cf.\ \cite[p. 148]{Smith2009DemocraticParticipation}: ``One of the main challenges facing internet-based engagement is the well-documented ‘digital divide’ that exists in terms of access to and proficiency in ICT.'' } In addition to significant differences in digital literacy \citep{Norris-2001,delborne-etal-2011}, it has been shown that large segments of the public are just not motivated to participate in online deliberation processes, cf.\ \citep[p.\ 366]{StrandbergGroenlund-2018-onlinedeliberation} and \citep{Smith2009DemocraticParticipation}. As a result, \emph{digital} minorities are not per se minorities offline. As long as the digital divide persists, we need to push back against taking brute numbers of approvals as indicative of the importance or quality of proposals. 
Moreover, by ensuring that also minorities’ proposals get heard, we may motivate them to become more active on DPs and thus to participate in democratic processes. Indeed, it has repeatedly been argued that to stimulate participation, it is important for participants that their efforts lead to a noticeable effect \citep{Stempeck2022GuidePlatforms}.

\subsection{Holistic Approaches to Approval-based Sorting}\label{subsec:twoalt}

In practice, DPs allow users to approve \emph{several} proposals during the online deliberation phase. This is so for a good reason: in line with the ideal of seeking consensus, DPs have the aim to generate a \emph{large}, well-worked-out option space, and within it, proposals that are supported by sufficiently large groups of users. Thus, approvals are often interpreted as expressing not only one’s favourite proposals, but also those proposals one is happy to endorse as part of a compromise. 

Given that users can approve several proposals, sorting merely by approval numbers poses an issue. Consider a very simple scenario with just 100 users and 30 proposals, all of which have been inspected by those users. Suppose that a group of 45 users approves proposals $x_1$ to $x_{10}$, a smaller (distinct) group of 35 users approves (distinct) proposals $y_1$ to $y_{10}$, and finally, 20 users approve proposals $z_1$ to $z_{10}$. Call these groups $A$, $B$, and $C$ respectively. If we sort them by approval numbers, proposals $x_1$ to $x_{10}$ will occupy the first ten positions in the list; followed by $y_1$ to $y_{10}$, and followed by $z_1$ to $z_{10}$. Thus, while only slightly less popular, proposals of type $y$ receive much less attention, and proposals of type $z$ hardly receive any attention – even if a significant part of the population approves (only) those proposals. In fact, while not even a majority of the population approves proposals of type $x$, these get absolute priority over all other proposals.

This scenario is artificial, but our argument applies \emph{a fortiori} to more realistic cases in which there are numerous proposals, numerous users, and some large group of users displays strong uniformity in approvals. In that case, the proposals supported by smaller groups are ``buried'' under the ones supported by the larger group. This phenomenon is well-known and has led to alternative sorting methods \citep{Behrens2014TheLiquidFeedback,Lackner2023Multi-WinnerPreferences}. 
Let us highlight two alternative approaches in broad terms:
\begin{itemize}
\item One can aim for a \emph{proportional sorting method}, i.e.\ generate a list that reflects the size of the different groups and the proposals they approve. In our example, this means that we still start with a proposal $x_i$, but this is followed by a proposal $y_i$ and then again a proposal $x_k$. However, relatively soon --- in the fifth position of the list --- we would have a proposal $z_k$ that represents group $C$.
\item One can aim for \emph{maximal coverage}, i.e.\ maximise the number of users that have \emph{some} proposal they approve high up in the list. In our example, this would mean that, although we will still put a proposal of type $x$ highest, this is followed by a proposal of type $y$, then one of type $z$. As a result, groups $A$, $B$, and $C$ are all represented in the the top three of proposals.
\end{itemize}
Table \ref{tab:example.rankings} illustrates these approaches. Note that the ideas of proportionality and maximal coverage are \emph{holistic}, in that they only make sense when applied to the list as a whole: in order to determine whether a given proposal $p$ should be sorted in position $n$, we cannot just compare this proposal to all other proposals one by one, but we have to take into account the \emph{set} of all proposals that were sorted on higher positions, and in light of that set, compare $p$ with all other remaining proposals. 

\begin{table}[ht]
\centering
\begin{tabular}{|>{\centering\arraybackslash}p{3cm}|
                >{\centering\arraybackslash}p{3cm}|
                >{\centering\arraybackslash}p{3cm}|}
\hline
\textbf{Max. Approvals} & \textbf{Proportional} & \textbf{Max. Coverage} \\
\hline
$x_1$ & $x_1$ & $x_1$\\
\hline
\multicolumn{1}{|c|}{$x_2$} & $y_1$ & $y_1$ \\
\hline
$x_3$ & $x_2$ & $z_1$ \\
\hline
\multicolumn{1}{|c|}{$x_4$} & $y_2$ & $x_2$ \\
\hline
$x_5$ & $z_1$ & $y_2$ \\
\hline
\multicolumn{1}{|c|}{$\vdots$} & \multicolumn{1}{c|}{$\vdots$} & \multicolumn{1}{c|}{$\vdots$} \\
\hline 
$x_{10}$ & $\vdots$ & $\vdots$ \\
\hline 
$y_{1}$ & $\vdots$ & $\vdots$ \\
\hline 
$\vdots$ & $\vdots$ & $\vdots$ \\
\hline 
$y_{10}$ & \multicolumn{1}{c|}{$\vdots$} & \multicolumn{1}{c|}{$\vdots$} \\
\hline
$z_1$ & \multicolumn{1}{c|}{$\vdots$} & \multicolumn{1}{c|}{$\vdots$} \\
\hline
\multicolumn{1}{|c|}{$\vdots$} & \multicolumn{1}{c|}{$\vdots$} & \multicolumn{1}{c|}{$\vdots$} \\
\hline
$z_{10}$ & \multicolumn{1}{c|}{$\vdots$} & \multicolumn{1}{c|}{$\vdots$} \\
\hline
\end{tabular}
\caption{Example lists for three different approaches.}
\label{tab:example.rankings}
\end{table}

Proportional sorting methods and methods that aim for maximal coverage have been studied mathematically in the field of Computational Social Choice. In particular, recent work on \emph{approval-based rankings} \citep{Skowron2017ProportionalRankings, Israel2021DynamicRankings} draws heavily on the theory of approval-based multiwinner voting \citep{Faliszewski2017Multiwinner, Faliszewski2017MultiwinnerTheory, Jaworski2022PhragmenProportionality, Lackner2017ConsistentRules, Lackner2023Multi-WinnerPreferences}. As shown there, if voters can approve several candidates and the aim is to select a winning committee (i.e.\ a \emph{set} of candidates of some pre-specified size), the very notion of proportionality can be cashed out in various specific, non-equivalent ways \citep{peters2020proportionality}. Some of these are computationally hard, while others can be implemented more easily. Similarly, depending on feasibility constraints, different algorithms for maximizing coverage can be used \citep{Thiele1895OmFlerfoldsvalg, Chamberlin1983RepresentativeRule, Procaccia2008OnRepresentation,Lu2011BudgetedMaking,Jaworski2022PhragmenProportionality}. While this is a rich and fascinating literature, what matters for our purposes is the basic idea behind both alternative approaches, as illustrated by the above example. 
 
In light of the above, there are different ways in which approvals can have a positive impact on the position of proposals. So which one would be best from the viewpoint of democratic theory?

Let us start with epistemic, instrumental arguments for deliberation. It is true that if some proposals received more approvals --- assuming that all proposals have been inspected equally well --- this may be taken as a sign of the quality of those proposals. However, this is only so if (a) every approval counts roughly equally and positively towards an estimation of a proposal's quality (i.e.\ all voters are equally competent, and better than random at guessing whether a proposal is good); (b) approvals simply add up (and thus, different voters evaluate any given proposal independently), and (c) the evaluation of each proposal is equally straightforward (thus, voters' competence does not vary across proposals). Each of (a)-(c) can however be questioned, in particular during the deliberation phase, where a broad range of very diverse proposals are still being discussed. As concerns (b), if proposals speak to certain groups in society, then it seems likely that the evaluation by members of those groups are strongly correlated — whether or not they are correct. As concerns (c), some proposals are typically more daring than others, and especially proposals that are more `out of the box' or less mainstream are typically harder to evaluate. So all in all, we believe that epistemic arguments do not allow us to single out any of the three approaches outlined above.\footnote{Maximizing coverage need not take \emph{absolute} priority over guaranteeing some minimal number of approvals for those proposals that end up high in the list. However, once a certain threshold of ``enough approvals'' has been passed, further differences between the number of approvals may well be outweighed by differences in the competence required to evaluate the proposals.}

When it comes to perceived legitimacy, the choice between sorting methods should ultimately be made on the basis of empirical evidence. One may expect that sorting by approval numbers has an advantage, since it is familiar from social media. However, in the context of participatory budgeting, it has been observed that proportional methods tend to be perceived as more legitimate than methods that choose the projects that received the highest number of approvals \citep{Yang-etal-DesigningDigitalVoting2024}. However, this kind of empirical work is still very rare and it remains to be seen whether it transfers to the choice of sorting methods for online deliberation. In any case, the perceived legitimacy of a given procedure hinges crucially on how we present and explain that procedure to those perceiving it, and thus on whether we \emph{can} justify the choice of the sorting method in independent terms, whether instrumental or intrinsic. In that sense, the discussion in this section would provide good reasons to think that holistic approaches can achieve good perceived legitimacy.

Let us then turn to more intrinsic arguments for deliberation.  
One crucial aspect we encountered here was that size should not matter: minorities should be heard just as well as majorities. This principle clearly favours an approach that maximises coverage. To be sure, by simply sorting the proposals approved by minorities high enough, we do not \emph{guarantee} that their concerns are taken into account. However, if the sorting method does have an impact on the attention proposals get, then the ideal of inclusiveness points towards maximizing coverage rather than sorting by approval numbers or proportionally. 

Let us make this slightly more concrete. Consider again our artificial example. Suppose that on average, users only really inspect the first three proposals they encounter.\footnote{Again, we make the example artificially small. In practice it may well be that they inspect far more, but then there may also be way more proposals than just 30.} Then if we sort by approval number, the proposals approved by group $C$ hardly get any attention, and thus this group is not properly included in the deliberation. These users have not been able to ``raise their concerns'', and users from groups $A$ and $B$ have not been able to benefit from their alternative perspectives. If instead we aim for maximal coverage, this problem does not arise.

Here again, the digital divide (cf.\ supra) further amplifies our argument. One may well take a diversity in approvals --- i.e. the fact that different proposals are proposed by distinct groups of users --- as a proxy for diversity of content and viewpoints. Thus, if one wants to promote the latter type of diversity, then again maximal coverage seems key.

\section{Concluding Remarks} \label{sec: conclusion}

In this paper we have argued for a principled approach to sorting methods. The key selling points of DPs are their low participation barrier and the possibility of asynchronic deliberation. However, these advantages create a challenge of managing and coordinating the ensuing stream of information. Using an ad hoc sorting method when showing proposals to users may result in arbitrary factors determining whose proposals get most attention, thereby risking to undermine the democratic ambitions of these platforms. We introduced a conceptual framework to compare sorting methods, studied existing DPs using this framework, and advanced two arguments against sorting by mere approval numbers and in favour of more integrated, holistic methods. 

We finish by listing a number of open problems for the science and practice of online deliberation, all of which relate to our discussion in this paper. 
\begin{itemize}
\item[Q1] What is the effect of sorting methods on the quality of proposals?
\item[Q2] What is the effect of sorting methods on the perceived legitimacy of online deliberation?
\item[Q3] What sorting methods do users themselves prefer or actually choose most in practice? 
\item[Q4] How can we inform users of the implications of a given sorting method?
\item[Q5] How can one measure the extent to which proposals have been inspected, and integrate this information in one's sorting method?
\end{itemize}
Following the current global trend in digitisation, the design and critical study of DDPs forms an integral part of democratic innovations. While sorting methods are only one aspect of this, we hope to have shown how already at this stage, basic democratic principles can inform design choices.

\bigskip

 \paragraph{Acknowledgements}
 We would like to express our gratitude to the following people for taking the time to answer our questions about their platform: Jan Behrens (LiquidFeedback -- cofounder), Janek Gulbis (Adhocracy+ \& Liquid Democracy e.V -- Project Coordination), Mathijs Kemp (Parta -- Partner \& Founder), Tietje Khieu (Adhocracy+ \& Liquid Democracy e.V -- Project Coordination), Anouk de Meulemeester (Go Vocal -- Government Success Manager), Andrés Pereira de Lucena (Decidim -- cofounder and lead developer), Simon Strohmenger (Consul -- Director), Ian Rieken (OpenStad \& Municipality of Amsterdam -- Software Developer). We are indebted to dr. Stefan Wintein, and the audiences at the Erasmus Institute of Philosophy and Economics research seminar (Erasmus University Rotterdam), the Action and Decision Theory Seminar (Munich Center for Mathematical Philosophy), the ALO meeting (Ghent University), the Social Epistemology of Argumentation conference (Vrije Universiteit Amsterdam) for comments on earlier versions of this paper. Finally, we sincerely thank the anonymous referees for their constructive feedback.

 \paragraph{Funding} This work was supported by the Nederlandse Organisatie voor Wetenschappelijk Onderzoek (NWO) VIDI project ENCODE – Explicating Norms of Collective Deliberation, grant number VI.Vidi.191.105.

\paragraph{Competing Interests} The authors have no relevant financial or non-financial interests to disclose.

\bigskip

\bibliography{References2}
\addcontentsline{toc}{chapter}{Bibliography}

\newpage

\appendix

\section{Number of Proposals per Case}\label{appsec:numbers}

\hide{}

\begin{table}[h!]
\centering
\small
\begin{tabularx}{\textwidth}{|l|l|l|X|}
\hline
\textbf{Platform} & \textbf{Location} & \textbf{\#Proposals} & \textbf{Source} \\ \hline

Go Vocal & Linz & 150 & {\tiny \href{https://www.govocal.com/casestudies/case-study-citizen-proposals-in-linz}{Go Vocal – Linz case}} \\ \hline
         & St. Louis & 3500 & {\tiny \href{https://www.govocal.com/case-studies/st-louis-collects-ideas-from-7000-residents-online}{Go Vocal – St. Louis case}} \\ \hline
         & Chile & 628 & {\tiny \href{https://www.govocal.com/case-studies/injuv-empowers-28-250-millennials-to-discuss-sustainable-development-ideas-for-chile}{Go Vocal – Chile case}} \\ \hline
         & Leuven & 2331 & {\tiny \href{https://www.govocal.com/case-studies/case-study-3000-citizens-contribute-to-leuven-multi-annual-plan}{Go Vocal – Leuven case}} \\ \hline
         & Ghent & 500 & {\tiny \href{https://www.govocal.com/case-studies/participatory-budgeting-ghent}{Go Vocal – Ghent case}} \\ \hline
         & Vienna & 1100 & {\tiny \href{https://www.govocal.com/case-studies/vienna-climate-team-award-winning-hybrid-engagement-project}{Go Vocal – Vienna case}} \\ \hline
         & Seattle & 1000 & {\tiny \href{https://www.govocal.com/case-studies/putting-inclusion-at-the-heart-of-seattles-comprehensive-plan}{Go Vocal – Seattle case}} \\ \hline

Consul   & Trier & 82 & {\tiny \href{https://docs.consuldemocracy.org/use_cases/germany/trier}{Consul – Trier case}} \\ \hline
         & San Pedro Garza Garcia & 900 & {\tiny \href{https://docs.consuldemocracy.org/use_cases/mexico/san-pedro-garza-garza}{Consul – San Pedro case}} \\ \hline
         & Glasgow & 301 & {\tiny \href{https://docs.consuldemocracy.org/use_cases/scotland/glasgow}{Consul – Glasgow case}} \\ \hline
         & Uruguay & 281 & {\tiny \href{https://docs.consuldemocracy.org/use_cases/uruguay/montevide}{Consul – Uruguay case}} \\ \hline

Decidim  & Europe & 18859 & {\tiny \href{https://decidim.org/blog/2022-06-01-the-conference-on-the-future-of-europe/}{Decidim – Europe case}} \\ \hline
         & Barcelona & 10860 & \citet[p.~25]{BarandiaranDecidimScience} \\ \hline
         & Brazil & 3500 & \citet[p.~25]{BarandiaranDecidimScience} \\ \hline

Your Priorities & Reykjavík & 6845 & {\tiny \href{https://citizens.is/portfolio_page/better_reykjavik/}{Your Priorities – Reykjavík}} \\ \hline
                & Iceland & 134 & {\tiny \href{https://citizens.is/portfolio_page/icelandic-constitution-crowdsourcing/}{Your Priorities – Iceland constitution}} \\ \hline
                & Scottish Parliament & 152 & {\tiny \href{https://citizens.is/portfolio_page/engage-scottish-parliament/}{Your Priorities – Scottish Parliament}} \\ \hline
                & Pirate Party & 553 & {\tiny \href{https://citizens.is/portfolio_page/piratenpartij-amsterdam/}{Your Priorities – Pirate Party}} \\ \hline
                & Pula & 83 & {\tiny \href{https://citizens.is/portfolio_page/pb-pula-croatia/}{Your Priorities – Pula}} \\ \hline

\end{tabularx}
\caption{A selection of cases by four different platforms, giving an indication of the number of proposals per case.}
\label{tab:proposal-numbers}
\end{table}

\section{Sorting Methods in Practice}\label{appsec:sortingmethods}

Table 3 
shows our findings on the use of sorting methods by the selected DPs. Note that for the ‘Evaluation → Approvals → Most’ column we include two related sorting options: one that orders proposals by the highest number of approvals, and another that ranks them by the best approval‑to‑disapproval ratio (as used, for example, on \platform{Adhocracy+} and \platform{Consul}). Because depending on the specific implementation, DPs sometimes do and sometimes do not support disapprovals, we present both options in one column. Furthermore, note that we were not able to find information on the default sorting method for Go Vocal. The ``combi'' column indicates that the platform uses a combination of factors in its sorting method.

\begin{landscape}
\scriptsize
\setlength{\tabcolsep}{3pt}
\renewcommand{\arraystretch}{1.3}

\begin{longtable}{|l|c|c|c|c|c|c|c|c|c|c|c|c|c|c|c|}
\hline
\textbf{Platform} 
& \multicolumn{3}{c|}{\textbf{Choice level}} 
& \multicolumn{3}{c|}{\textbf{Content}} 
& \multicolumn{4}{c|}{\textbf{Metadata}} 
& \multicolumn{3}{c|}{\textbf{Evaluation}} 
& \textbf{Random} 
& \textbf{Combi} \\
\cline{2-16}
& \textbf{Dev.} & \textbf{Host} & \textbf{User}
& \multicolumn{2}{c|}{\textbf{Cost}} & \multirow{3}{*}{\textbf{Tags}} 
& \multicolumn{2}{c|}{\textbf{Date}} & \multirow{3}{*}{\begin{tabular}{@{}c@{}}Most\\comments\end{tabular}} & \multirow{3}{*}{\begin{tabular}{@{}c@{}}Most\\authors\end{tabular}} 
& \multicolumn{3}{c|}{\textbf{Approvals}} 
& \multirow{3}{*}{\textbf{ }} 
& \multirow{3}{*}{\textbf{ }} \\
\cline{5-6} \cline{8-9} \cline{12-14}
& & & 
& High & Low & 
& New & Old & & 
& Most & Least & \begin{tabular}{@{}c@{}}Propor-\\tional\end{tabular} & & \\
\hline
\platform{Adhocracy+}      & x  & x  & x &   &   & x & \color{purple}d  &   & x &   & x &   &   & &  \\
\platform{Consul}          & x  & x  & x & x &   & x & x &   &   &   & x &   &   & x & \color{purple}d\\
\platform{Decidim}         & x  & x  & x  & x & x & x  & x &   & x & x & x  &   &  & \color{purple}d &  \\
\platform{Go Vocal}        & x  & x & x &   &   & x & x & x &   &   & x &   &   & x & x  \\
\platform{LiquidFeedback}  & x &   &   &   &   &   &   &   &   &   &   &   & \color{purple}d  &   & \\
\platform{OpenStad}        & x  & x & x & x & x & x  & x & x &   &   & x & x &   &   & \\
\platform{Parta}           & x &   &  &   &   &   &   &   &   &   &   &   &   & \color{purple}d &  \\
\platform{Your Priorities} & x  & x  & x &   &   &   & \color{purple}d & x & x &   & x &   &   & x & \\
\hline
\caption{Overview of sorting methods for online deliberation used by DPs. The default sorting method is denoted by ‘\color{purple}d\color{black}'.}
\end{longtable}
\label{tab:overview}
\end{landscape}

\end{document}